\documentclass{elsarticle}

\usepackage{epsfig}
\usepackage{amssymb}
\usepackage{amsmath}

\begin{document}

\begin{frontmatter}

\title{Study of the 1D anisotropic Kondo necklace model at criticality via an entanglement entropy estimator}

\author{A. Saguia\corref{sag}}
\ead{amen@if.uff.br}
\address{Instituto de F\'{\i}sica - Universidade Federal Fluminense, Av. Gal. Milton Tavares de Souza, s/no,  Niter\'oi, 24210-346, RJ - Brazil}
\cortext[sag]{}

\begin{abstract}

We use an estimator of quantum criticality based on the entanglement entropy to discuss the ground state properties 
of the 1D anisotropic Kondo necklace model. 
We found that the  $T=0$ phase diagram of the model is 
described by a critical line separating an antiferromagnetic phase 
from a Kondo singlet state. Moreover we calculate the conformal anomaly on the critical line and obtained that 
$c$ tends to $0.5$ as the thermodynamic limit is reached. Hence we conclude that these transitions belong to Ising universality 
class being, therefore, second order transitions instead of infinite order as claimed before.

\end{abstract}

\end{frontmatter}

\section{Introduction}

Heavy-fermion compounds belong to a class of materials which exhibit many intriguing and anomalous phenomena as, for example, the Kondo effect.
In these materials the $f$ shell electrons present an unstable character that oscillates between localized and itinerant. 
Because of this ambiguous behaviour they can be found in a broad variety of states including metallic, 
superconducting, insulating and magnetic states~\cite{1,2}. Interestingly, it has been showed that most of the properties of these systems
can be attributed to their proximity to a magnetic quantum critical point (QCP) \cite{3}, and
due to this important characteristic, the study of these system at low temperatures has been of great interest recently (see, for example, Refs.~\cite{n1,n2}).

A standard Hamiltonian which has been largely used to describe the physical properties of these compounds
is the Kondo lattice model \cite{kl,don}. 
It assumes the presence of one localized impurity spin
on each site of the lattice, coupled to the conduction electrons of the metal. The interaction between the magnetic moments of the $f$ electrons of the impurities
and those of the conduction electrons, $J$, is responsible for the Kondo effect. This Kondo interaction tends to compensate the local moments, 
forming singlets, giving rise to a non 
magnetic ground state. At the same time, because of the large concentration of impurities, it appears an interaction between their magnetic moments. This is an inter-site
coupling of the Ruderman-Kittel-Kasuya-Yosida (RKKY) type and it is mediated by the conduction electrons. The RKKY interaction
tends to establish a long range magnetic order in the system, in general, an antiferromagnetic ground state.
This competition induces a magnetic quantum phase transition in this model which is the actual case in most heavy
fermions materials.
A simplified version of the Kondo lattice Hamiltonian was introduced by Doniach in Ref.~\cite{don}. This model, called Kondo necklace (KN), replaces, into the
Kondo interaction, the spins of the conduction electrons by a set of pseudo-spins on a linear lattice, the charge degrees of freedom being frozen
out. In spite of this approximation, the interplay between the Kondo mechanism and magnetic ordering remains as an essential feature.
Although the KN model has been extensively studied over the last decades many questions about the
phase diagram and magnetic properties of intricate versions of this model still remain open and we believe it will still be necessary some effort to be completely 
understood.

Among the recently studied systems is the one-dimensional KN model in the
presence of an Ising-like anisotropy. The general Hamiltonian that describes the system is given by
\begin{equation}
H = \sum_{i=1}^{N-1} W(\sigma_{i}^{x}\sigma_{i+1}^{x} + (1-\delta)\sigma_{i}^{y}\sigma_{i+1}^{y}) +\sum_{i=1}^{N} J\vec{S}_{i}.\vec{\sigma}_{i},
\label{eqh}
\end{equation}
where $\sigma^{\mu}$ and  $S^{\mu}$, $\mu = x,y,z$, are spin-1/2 Pauli matrices denoting the spin of the conduction electrons and those of the 
local moments, respectively.
$J$ is the intra-site exchange interaction between them. The indices
$i$ and $i+1$ denote nearest neighbors on a chain of N sites and $W$ is an antiferromagnetic coupling which represents the hopping 
of the conduction electrons between neighborings sites (see in Fig.~\ref{f1} a schematic representation of the chain with $N=6$ sites, i.e, $12$ spins).
The Ising like anisotropy parameter ($\delta$) varies from zero to one.  

The zero temperature phase diagram of this model in the ($\delta$,$K$) plane, with $K=J/W$, has been studied
by different methods as renormalization group~\cite{saguia}, Lanczos~\cite{langari}, and DMRG~\cite{colomb2}. 
The main results found in the literature can be summarized as follows.
The line $K = 0$ corresponds to the one-dimensional (1D), purely anisotropic XY model
which is ordered at $T = 0$~\cite{bc}.
For $K\neq0$ and $\delta=0$ case, the original KN model is recovered and it is already well established that any finite value of the interaction 
$J$ give rise to a non-magnetic Kondo state~\cite{rap,mouk,scal}. 
In the full anisotropic case ($\delta=1$), there is an unstable fixed point separating an antiferromagnetic phase for small 
values of $K$ from a spin compensated, Kondo-like phase, reached in the strong $J$ interaction regime. 
The exact value of $K_{c}$  is not known but there is a consensus that this transition is in the same class of universality of 
the 1D quantum Ising model~\cite{rap}. The great controversy arises when we consider the anisotropy parameter in the region $0<\delta<1$. 
In this case the two phases described above are still present but it is on debate if   
a critical value of anisotropy is required for the appearance of long-range magnetic order, as predicted in Ref.~\cite{saguia}, or if it is 
present for any value of anisotropy, as reported in Refs.~\cite{langari,colomb2}. Other important point of discussion is the class of universality
of the ordered-disordered transition. RG calculation indicates it is a second order phase transition but
the authors of Ref.~\cite{colomb2} proposed it is of Kosterlitz-Thouless type. 

To investigate the critical behaviour of the 1D anisotropic KN model at zero temperature 
we use an estimator of quantum criticality based on the behaviour of the entanglement entropy in gapless and gapfull systems. 
This method has proved to be a powerful numerical tool to
precisely locate quantum critical points and calculate the central charge with low computational cost (small lattice sizes) 
in a large variety of 1D quantum systems~\cite{ca,nish}.
By considering systems with sizes of up to 24 spins,
we found 
that the T=0 phase diagram of this model is described by a critical line separating an antiferromagnetic long range order, 
which is present for any finite value of anisotropy, from a non-magnetic Kondo singlet phase. 
This general result is in accordance to  Lanczos and DMRG calculations. 
However, we have calculated the conformal anomaly on the critical line and obtained that 
$c$ tends to $0.5$ as the system size increases. Hence we conclude that the transitions 
for any $\delta\neq0$ belong to Ising universality class and, therefore, they are of the second order kind instead of 
infinite order as claimed in Ref.~\cite{colomb2}.

The paper is organized as follows: In the next section we outline the formalism adopted and, in the last
section, we present and discuss our results.
  
\section{Formalism}

In this work we present a systematic study of the quantum behaviour of the model described by Eq.~\ref{eqh} at $T=0$. To identify the critical
coupling ($K=J/W$) and anisotropy ($\delta$) separating the various quantal phases of the system 
we employ an estimator of quantum criticality based on the von Neumann entanglement entropy. 
It works in the following way.
Let us consider a quantum system of $L$ spins in a pure state $|\psi\rangle$ 
and a bipartition of the system into two blocks: a block of $l$ contiguous spins and other containing the spins of the
rest of the chain ($L-l$ spins). The entanglement entropy between 
$l$ and $L-l$ is given by
\begin{equation}
{\cal S}(L,l)=-\textrm{Tr} \left( \rho_l \ln \rho_l \right) 
\label{vn}
\end{equation}
where $\rho_l=\textrm{Tr}_{L-l} \rho$ denote the reduced 
density matrix for block $l$, with $\rho=|\psi\rangle\langle \psi|$. 
As it has been shown in Refs.~\cite{ent1,ent2,ent3,ent4,ent5} the quantity defined in Eq.~\ref{vn} presents a very
interesting scaling behaviour in 1D systems. Suppose this system is dependent of a given parameter, let us say $\lambda$, and at
$\lambda=\lambda_c$ it presents a QCP.  In the critical point, 
conformal invariance implies a diverging logarithmic scaling which can be written as
\begin{equation}
{\cal S}(L,l) = \gamma \ln[\frac{L}{\pi}\sin(\frac{\pi l}{L})] + \beta
\label{vns}
\end{equation}
where $\beta$ is a nonuniversal constant and $\gamma$ is a constant related to the central charge, namely $\gamma = c/3$ when 
periodic boundary conditions are adopted in the chain. 
If $\lambda\neq\lambda_c$ and $(l,L-l)\to\infty$, then the entanglement entropy $S(L,l)$ is not only independent of $l$ but also independent of $L$. There is entanglement only between sites separated by a distance of the order of the correlation length ($\xi$), and this quantity of entanglement is unchanged by a variation of either $L$ or $l$ when both $l$ and $L-l$ are very large.

The estimator used throughout the paper was proposed in Refs.~\cite{ca,nish} and it is given by the difference of the entanglement entropy obtained for two subsystems of
different sizes, for example, $l$ and $l^{\prime}$, of a system of total size $L$. We write this as: 
\begin{equation}
\Delta {\cal S} = {\cal S}(L,l) - {\cal S}(L,l^{\prime}),  
\end{equation}
where ${\cal S}(L,l)$ and ${\cal S}(L,l^{\prime})$ are the entanglement entropies between $l$ and $(L-l)$ and between $l^{\prime}$ and $(L-l^{\prime})$, respectively.
From the expressions above, we see that as $(L,l,l^{\prime}) \rightarrow \infty$, $\Delta {\cal S}\neq 0$
at the critical point while it is zero for any value of $\lambda\neq\lambda_c$. Therefore $\Delta {\cal S}(\lambda)$ is a good indicator
of quantum phase transition in the thermodynamic limit. For
a finite size system the scaling of ${\cal S}(L,l)$ described above is not exact anymore but the result obtained 
for $\Delta {\cal S}$ can be easily generalized. In this case, it is expected that 
$\Delta {\cal S}\neq 0$ for all values of $\lambda$ but
at $\lambda=\lambda_c$ it attains its maximum value. In this way, $\Delta {\cal S}$ works as an estimator of criticality:
to locate a QCP for a fixed finite system size we should look for the value of $\lambda$ 
for which $\Delta {\cal S}$ is a maximum. As L increases the peak around $\lambda_c$ should become narrower, so that in the thermodynamic limit it is
the only value different from zero. It is important to emphasize that 
the choice of $l$ and $l^{\prime}$ for this method is arbitrary but the finite-size effects on  ${\cal S}(L,l)$ are smaller if 
they are chosen around the middle of the chain.

An interesting characteristic of this method is that it allows to obtain a unique value of the critical parameter and  central charge for each $L$,
enabling us to estimate their values in the thermodynamic limit by extrapolation. This approach has been tested in different models and the results have shown that this is an excellent numerical method to study quantum criticality in 1D systems~\cite{ca,nish}.

\section{Results and discussion}

In order to apply the estimator to the KN model we consider subsystems containing $l$ and $l^{\prime}$ spins of a periodic chain with $L$ spins. 
Note that, in Eq.~\ref{eqh}, N represents the number of sites of the system in an open chain, since there are two spins at each site it implies that $L=2N$
(As an example, we show in Fig.(1), a schematic representation of the chain with $L=12$ and $l^{\prime}=4$).
The adoption of periodic boundary condition in the KN model means that we should take the first sum in Eq.~\ref{eqh} up to $N$, considering $\sigma_{N+1} = \sigma_{1}$, keeping the second one as it is.  
\begin{figure}[h]  
\centering{\includegraphics[angle=0,scale=0.32]{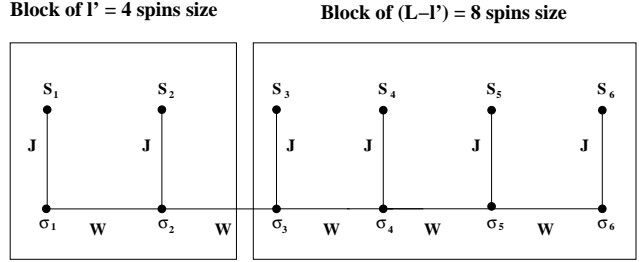}}  
\caption{ \label{f1} Schematic arrangement of the KN chain
with $L=12$ spins.  The spin-1/2 operators $\sigma^{\mu}$ and $S^{\mu}$,
where $\mu=x,y,z$, denote the conduction electrons and the spins of the
local moments, respectively. We also show the subsystem of size $l^{\prime}=4$ spins and its complementary block.}
\end{figure}

Initially, to test the method, we will discuss the critical point and the central charge of the KN model for the specific case of $\delta =1$. It is a good starting-point since
there is not a great controversy about these results in the literature.
To calculate $\Delta {\cal S}$ we can choose the blocks, $l$ and $l^{\prime}$, of arbitrary size. In this primary example, we will use two different bipartitions of the system: 
(a) $l=L/2$ and $l^{\prime}=L/4$ and (b) $l=L/2$ and $l^{\prime}=L/2-2$. In the first case we take $l$ and $l^{\prime}$ fixed ratios of $L$ such that 
the difference between them, $l-l^{\prime}=L/4$, increases
with $L$, for the second bipartition, this difference remains constant, $l-l^{\prime}=2$. 
We will compare the accuracy of the numerical results and the optimal bipartition will be used to investigate 
the critical line in the interval $0<\delta<1$.  

The basic steps of the approximation, valid for the two bipartitions, are: firstly, we use power method (an iterative eigenvalue algorithm~\cite{calcnum}) to obtain the ground state
of the chain for a given fixed value of $L$ up to $24$ spins. Then, we calculate 
${\cal S}(L,l)$ and ${\cal S}(L,l^{\prime})$ as defined in Eq.~\ref{vn} and 
subtract them to obtain $\Delta {\cal S}$. 
To locate the QCPs we fix the value of $\delta$=1 and look for the value of $K$ 
that gives the maximum value of $\Delta {\cal S}$. It is important to emphasize that, due to the geometry of the lattice, we just can choose
some specific values of $L$ to study the correlations along the chain. In fact, the values allowed for $L$ depend on the bipartition of the system: 
for the case (a), the chain should
contain $L=8,16$, and $24$ spins, while for (b), $L$ can be equal to $12,16,20$, and $24$ spins.
The critical coupling in the thermodynamic limit is
estimated, in both cases, by extrapolating  the values of $K$ found for each spin system size, for larger $L$.
Let us present the results for the two bipartitions separately.

(a) $l=L/2$ and $l^{\prime}=L/4$

As discussed in the last section, as $(l,l^{\prime},L)$ increases, $\Delta {\cal S} \rightarrow 0$, except in the critical point where it tends to:
\begin{equation}
\Delta {\cal S} = {\cal S}(L,\frac{L}{2}) - {\cal S}(L,\frac{L}{4}) 
= \frac{\gamma}{2}\ln(2).
\end{equation}
This result can also be written as:
\begin{equation}
c = \frac{6.\Delta {\cal S}}{\ln(2)}.
\label{nc}
\end{equation} 
From this expression we see that, just like $\Delta {\cal S}$, $c$ is a maximum at the critical point. 
Thus, once we have localized  $K_c$ we use the corresponding maximum value of $\Delta {\cal S}$ to calculate $c$ in Eq.~\ref{nc} for each L.
The value of $c$ in the thermodynamic limit is estimated by extrapolation. 

\begin{figure}[h!]
\centering {\includegraphics[scale=0.3]{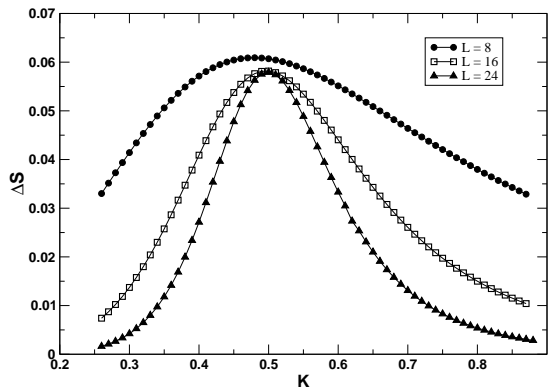}}
\caption{Dependence of the difference of entropy $\Delta {\cal S}$ with $K$ for various values of L, taking the bipartition as defined in (a). 
See that, as L increases, the maximum value of $\Delta {\cal S}$ tends to $K = 0.5$ and   
the peak around $K_c$ become narrower. These characteristics lead us to conclude that $K = 0.5$ 
corresponds to the QCP of model for $\delta=1$.}
\label{nf2}
\end{figure}

In Fig.~\ref{nf2} we show the evolution of
$\Delta {\cal S}$ as a function of $K$ for $\delta=1$. As we can see, the maximum of $\Delta {\cal S}$ occurs in 
$K=0.5$ and, as $L$ increases, the peak becomes narrower around this value. With the value of $K_c$ in hands, the calculation
of the central charge is a simple application of Eq.~\ref{nc}. 
In Table~\ref{ntab2} we show the values of $K_c$ and $c$ as a function of $L$. 
See that theses parameters converge fastly to the values $K_c=0.5$ and $c=0.5$, as $L$ increases. Therefore we believe that they
are good numerical approximation to the true values of the quantum critical point, which are reached only in the thermodynamic limit.
Based on this approximation, we conclude that $K_c=0.5$ corresponds to the QCP of the system for $\delta=1$ and that, 
this transition, characterized by $c=0.5$, belongs to the ising universality class. 
\begin{table}[h]
\centering
\begin{tabular}{c|ccc}
$L$ & $K_{c}$ & & $c$ \tabularnewline
\hline
\hline 
8  & 0.48  & &  0.5273\tabularnewline
16  & 0.50  & &  0.5035\tabularnewline
24  & 0.50  & &  0.5013\tabularnewline
\hline 
Estimate to $L\rightarrow\infty$  & 0.50  & & 0.5\tabularnewline
\end{tabular}
\caption{System size dependence of the critical coupling, $K_{c}$, and the central charge, $c$, 
at the QCP of the anisotropic KN model for $\delta =1$, by
considering the bipartition (a). The values of these parameters in the thermodynamic limit are estimated by extrapolation.}
\label{ntab2}
\end{table}

\noindent b) $l=L/2$ and $l^{\prime}=L/2 - 2$

By using this bipartition we obtain that, as $(l,l^{\prime},L)$ increases:
\begin{equation}
\Delta {\cal S} = {\cal S}(L,\frac{L}{2}) - {\cal S}(L,\frac{L}{2}-2)= 
\begin{cases}
-\gamma\ln[\cos(\frac{2\pi}{L})], & K=K_{c} \\
0,  & K\ne K_{c}
\end{cases}
\end{equation}
From this expression we can see that, for $K=K_{c}$, $\Delta {\cal S}\sim L^{-2}$, which tends to zero as $L\to\infty$. 
However, we should point out that $\Delta {\cal S}$ 
is still a maximum at the critical point because the vanishing of $\Delta {\cal S}$ for $K\ne K_{c}$ is much faster (exponential) than for $K=K_{c}$ (power law). 
Therefore, $\Delta {\cal S}$ still works as an estimator of quantum criticality and the method can be used as before.
The central charge at the critical point can be obtained by:
\begin{equation}
c = \frac{-3 \Delta {\cal S}}{\ln(\cos(2\pi/L))}.
\label{c}
\end{equation}

\begin{figure}[h!]
\centering {\includegraphics[scale=0.3]{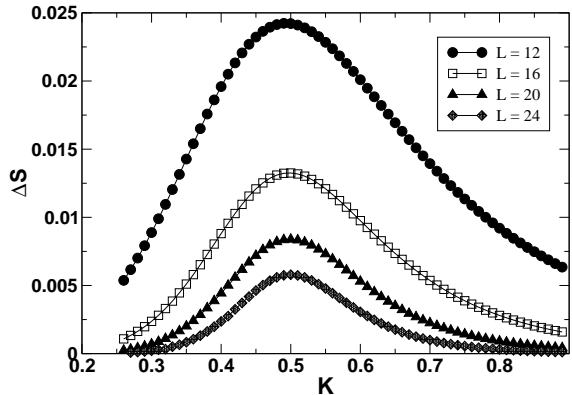}}
\caption{Dependence of the difference of entropy $\Delta {\cal S}$ with $K$ for various values of L, by considering the bipartition (b). 
See that, although the height of $\Delta {\cal S}^{max}$ decreaseas as $L$ increases, $\Delta {\cal S}$ presents a peak increasingly narrow around $K=0.5$
indicating that this is the critical point in the model for $\delta=1$.}
\label{f2}
\end{figure}

In Fig.~\ref{f2} we show the evolution of $\Delta {\cal S}$ as a function of $K$ for $\delta=1$. As already expected, the height of 
$\Delta {\cal S}^{max}$ decreases as $L$ increases. However, as discussed above, the position of the maximum value of $\Delta {\cal S}$ still 
indicates the critical point. As it can be seen, $\Delta {\cal S}$ presents a peak increasingly narrow around $K=0.5$ as $L$ increases. 
To calculate the central charge and discuss the class of universality of the transition we use Eq.~\ref{c}.
We show in Table~\ref{tab2} the finite size values of $K_c$ and $c$ as a function of $L$. See that, as $L$ increases, $c\rightarrow 0.5$. 
The fast convergence of $c$ leads to the conclusion that 
this is a good approximation to the value of the central charge in the thermodynamic limit for the model with $\delta = 1$. 

\begin{table}[h]
\centering
\begin{tabular}{c|ccc}
$L$ & $K_{c}$ & & $c$ \tabularnewline
\hline
\hline 
12  & 0.49  & &  0.5054\tabularnewline
16  & 0.50  & &  0.5023\tabularnewline
20  & 0.50  & &  0.5011\tabularnewline
24  & 0.50  & &  0.5009\tabularnewline
\hline 
Estimate to $L\rightarrow\infty$  & 0.50  & & 0.5\tabularnewline
\end{tabular}
\caption{System size dependence of the critical coupling $K_{c}$ and the central charge $c$ at the QCP of the anisotropic KN model for $\delta =1$, 
by considering the bipartition (b).
The values of these parameters in the thermodynamic limit are estimated by extrapolation.}
\label{tab2}
\end{table}

Based on the results showed above we can conclude that the two bipartitions used in this first analysis are equivalent and that 
there is not a real advantage in one choice over the other. 
Independently of the bipartition used, the estimated value for $K_c$ and $c$, in the thermodynamic limit, are the same. Moreover, in both 
cases, $c$ is expressed with high precision (the uncertainty is in the third decimal place)
even in a very small chain (just 16 spins). Therefore, as the results seem to be not affected by the choice of one of the bipartitions, 
we will choose to continue our investigation by taking only the bipartition (b) from now on.
It is worth mentioning that, in using (a) for our model, $L$ can assume only three values and the first one is $8$, 
which is very small and does not produce very precise results. In (b), $L$ can be equal to $12,16,20$ and $24$ spins which is more
convenient for the finite size analysis.    

The whole process is then repeated for other values of anisotropy $\delta$, so that we obtain the critical points as showed in Table~\ref{tab1}.
These results can be summarized in the $K$ vs $\delta$ phase diagram showed in Fig.~\ref{f1}. For comparison we also show in this figure 
the results obtained in Refs.~\cite{saguia,langari,colomb2}.
As it can been seen, our calculations indicate that the system exhibits two phases for any finite value of anisotropy. A critical line separates the antiferromagnetic phase 
which is present for small values of $K$, from a spin compensated, Kondo-like phase, reached 
in the strong $J$ interaction regime. Therefore we conclude that, on the contrary of proposed in Ref.~\cite{saguia}, the long range order is
always present in the KN model independently of the value of anisotropy. Note that our results agree qualitatively to those 
obtained via Lanczos and DMRG method. 
\begin{table}[h]
\centering
\begin{tabular}{c|ccccccccccc}
$\delta$ & 0.0 & & 0.1 & & 0.3 & & 0.5 & & 0.7 & & 1.0 \tabularnewline
\hline
$K_c$    & 0.0 & & 0.40& & 0.46 & & 0.48 & & 0.49 & & 0.50 \tabularnewline
\hline 
\end{tabular}
\caption{The critical coupling, $K_{c}$, of the KN model for different values anisotropy ($\delta$).}
\label{tab1}
\end{table}
\begin{figure}[h!]
\centering {\includegraphics[scale=0.3]{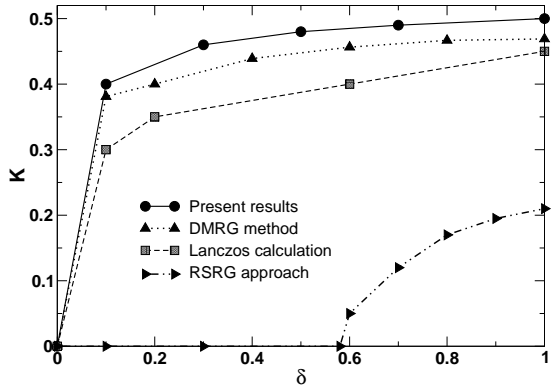}}
\caption{Phase diagram of the anisotropic KN model obtained via criticality estimator in chains of size up to 24 spins). 
The critical line separates the antiferromagnetic phase (below each line) from the non-magnetic ground state.
For comparison we also show the results obtained by DMRG method~\cite{colomb2}, 
Lanczos calculation~\cite{langari}, and RSRG approach~\cite{saguia}.}
\label{f1}
\end{figure}

\begin{figure}[h]
\centering {\includegraphics[scale=0.3]{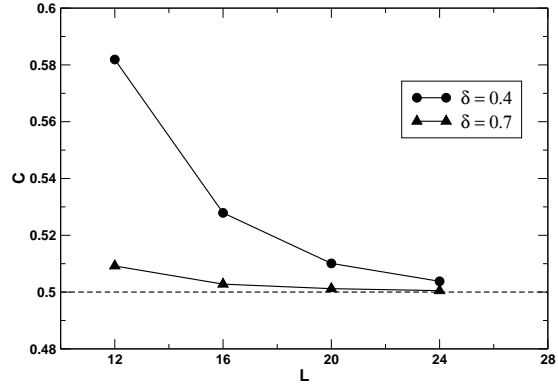}}
\caption{Dependence of the central charge with the system size at the critical points of $\delta=0.4$ and $\delta=0.7$. The dotted line denotes the value $c=0.5$ corresponding
to estimate of $c$ in the $L\rightarrow\infty$ limit.}
\label{fig3}
\end{figure}

Following the procedure described above, we have calculated the central 
charge at the critical point for several values of $\delta$. Our results show that the estimated value for $c$ is $0.5$ for the entire line. As an example of our calculations, 
we show in Fig.~\ref{fig3} the values of $c$ as a function of $L$ obtained
for $\delta=0.4$ and $\delta=0.7$. We see clearly that  $c$ tends to $0.5$, as L increases, for both values of anisotropy.
Therefore, we conclude that the transitions along the line belong to the quantum 1D Ising model universality class and they are of the second order type. 

\section{Summary and Conclusion}

In summary, we have examined the QPTs of the 1D Kondo necklace model
in the presence of an Ising-like anisotropy. 
The model is suitable to describe heavy
fermion systems and emphasizes magnetic degrees of freedom neglecting charge fluctuations. 
By using an estimator of quantum criticality based on the behaviour of the entanglement entropy~\cite{ca,nish},
we found that the T=0 phase diagram of this model is described by a critical line separating an antiferromagnetic long range order, 
which is present for any finite value of anisotropy, from a non-magnetic Kondo singlet phase. 
This general result is in accordance to  Lanczos and DMRG calculation. 
However we have calculated the central charge on the critical line and obtained that 
$c \rightarrow 0.5$, as $L$ increases, for any $\delta\neq0$. Hence we conclude that these transitions 
along the line belong to the Ising universality class and, therefore, they are of the second order type instead of 
infinite order as claimed in Ref.~\cite{colomb2}.

It is very interesting to note the similarity between the phases diagrams of the anisotropic KN model, as obtained here, and that of the $XY$ model in a transverse 
field ($\lambda$) obtained by Barouch and McCoy in Ref. \cite{bc}. 
The isotropic ($\delta=0$) version of both systems presents an infinite order critical point ($J=0$ for the KN model and $\lambda=1$ for the $XY$ Hamiltonian), however,
any perturbation in the anisotropic parameter (any $\delta>0$) turns the critical behaviour of the spin chains from the universality class of the $XX$ model into that of the Ising model. In the KN model, the $RKKY$ ($J$) interaction plays the role of the transverse field in the $XY$ model: 
it destroys the correlations along the chain leading the system to a paramagnetic phase. We leave this point to be further explored in future works.

\end{document}